\documentclass[journal]{vgtc}                

\usepackage{microtype}                 
\PassOptionsToPackage{warn}{textcomp}  
\usepackage{textcomp}                  
\usepackage{mathptmx}                  
\usepackage{times}                     
\usepackage{cite}                      
\usepackage{tabu}                      
\usepackage{booktabs}                  
\usepackage{amsmath}

\usepackage{placeins}

\usepackage{graphicx}
\graphicspath{ {images/} }

\onlineid{0}

\vgtccategory{Research}
\vgtcpapertype{application/design study}

\newcommand{\trans}[2]{^\textrm{#1}\mathbf{T}_\textrm{#2}}

\title{Augment Yourself: Mixed Reality Self-Augmentation Using Optical See-through Head-mounted Displays and Physical Mirrors}

\author{Mathias Unberath, Kevin Yu, Roghayeh Barmaki, Alex Johnson, and Nassir Navab}
\authorfooter{
\item M. Unberath and K. Yu have contributed equally.
\item
 M. Unberath, K. Yu, R. Barmaki, A. Johnson, and N. Navab were with Johns Hopkins University, Baltimore, MD.
\item
 A. Johnson was with Johns Hopkins Hospital, Baltimore, MD.
 \item Send correspondence to \url{unberath@jhu.edu}.
}

\shortauthortitle{Unberath and Yu \MakeLowercase{\textit{et al.}}: Augment Yourself}

\abstract{
Optical see-though head-mounted displays (OST HMDs) are one of the key technologies for merging virtual objects and physical scenes to provide an immersive mixed reality (MR) environment to its user. A fundamental limitation of HMDs is, that the user itself cannot be augmented conveniently as, in casual posture, only the distal upper extremities are within the field of view of the HMD. Consequently, most MR applications that are centered around the user, such as virtual dressing rooms or learning of body movements, cannot be realized with HMDs.\\
In this paper, we propose a novel concept and prototype system that combines OST HMDs and physical mirrors to enable self-augmentation and provide an immersive MR environment centered around the user. Our system, to the best of our knowledge the first of its kind, estimates the user's pose in the virtual image generated by the mirror using an RGBD camera attached to the HMD and anchors virtual objects to the reflection rather than the user directly.\\
We evaluate our system quantitatively with respect to calibration accuracy and infrared signal degradation effects due to the mirror, and show its potential in applications where large mirrors are already an integral part of the facility. Particularly, we demonstrate its use for virtual fitting rooms, gaming applications, anatomy learning, and personal fitness.\\
In contrast to competing devices such as LCD-equipped smart mirrors, the proposed system consists of only an HMD with RGBD camera and, thus, does not require a prepared environment making it very flexible and generic. In future work, we will aim to investigate how the system can be optimally used for physical rehabilitation and personal training as a promising application.
}


\vgtcinsertpkg

\begin{document}

\firstsection{Introduction}
\maketitle

\label{intro}
\subsection{Background}

\begin{figure}[t!]
\centering
\includegraphics[width = 0.95\linewidth]{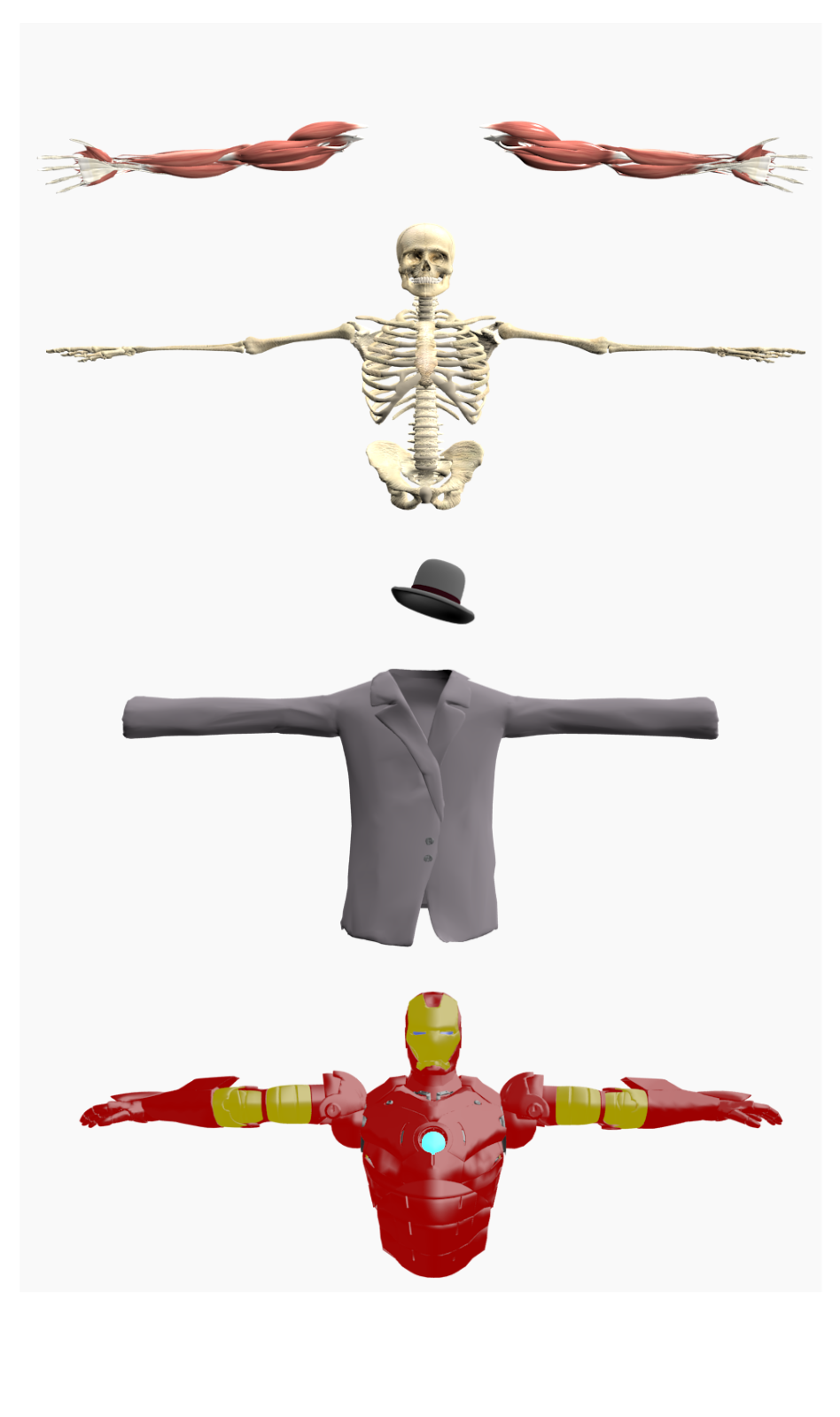}
\caption{Listed from top to bottom, we show virtual models that could be used in personal fitness (arm muscles), anatomy learning (skeleton), virtual dressing rooms (jacket and hat), and entertainment (Ironman suit), respectively.}
\label{Figure:VRModels}
\end{figure}

Mixed reality (MR) is receiving increasing attention in many different applications ranging from entertainment~\cite{ma2011serious,aukstakalnis2016practical,Infrared2017}, education~\cite{lee2012augmented,wu2013current,Bork2017}, to medicine~\cite{shuhaiber2004augmented,fischer2016preclinical,Bork2017,qian2017technical}. It is considered as one of the strategic technological trends~\cite{avila2016augment,schmalstieg2016augmented}. In the MR context, optical see-through head-mounted displays (OST HMDs) are particularly promising as they are mobile and enable merging of virtual views with physical scenes~\cite{liu2008optical} providing a fully immersive MR environment to the user. This positive development is propelled forward as more and more consumer grade devices, such as the Microsoft HoloLens (Microsoft Corporation, Redmond, WA) or the Meta 2 (Meta Company, San Mateo, CA), become available to the general public.\\
One of the most straightforward concepts of MR is the augmentation of real, physical items by placing virtual objects such that they remain in a certain spatial relation to them, for example a virtual vase on top of a real table. This type of augmentation is enabled by simultaneous localization and mapping (SLAM) algorithms~\cite{durrant2006slam,Holmdahl2015,Microsoft2017coord,kress2017hololens}. Based on the depth and RGB sensor data, the HMD incrementally generates a spatial map of its surrounding, local environment while continuously estimating its pose therein. Therefore, virtual objects can be placed with respect to the spatial map (also referred to as \emph{anchoring}) that remains largely constant over time and are only displayed when they enter the field of view of the HMD. Although some aspects of anchoring virtual objects to a static scene are still subject to ongoing research, such as the interaction of virtual and real objects~\cite{nuernberger2016snaptoreality}, or the estimation of lighting conditions for realistic rendering~\cite{richter2016instant}, the most crucial problem of SLAM, despite room for improvement, can be considered well addressed.\\
Aforementioned concept of positioning virtual objects in a certain spatial relation to the physical world can be extended rather straightforwardly to anchoring to dynamic objects, such as humans. In this case, more sophisticated tracking solutions are required that employ dense or sparse methods, such as deformable SLAM~\cite{newcombe2015dynamicfusion,innmann2016volumedeform} or human skeleton tracking~\cite{zhang2012microsoft,zhou2016sparseness}. Methods devised for human skeleton tracking paired with SLAM and OST HMDs allow for the augmentation of persons in the local environment, for instance by anchoring a deformable avatar to their skeleton.\\
Despite the impressive performance of modern OST HMDs that, paired with novel tracking algorithms, provide convincing MR environments, a fundamental problem remains unaddressed: the user wearing the OST HMD himself, cannot conveniently be augmented. The reason seems straightforward: at usual, relaxed head poses and body postures the user does not occupy much of the HMD's field of view, only the distal upper extremities (forearms and hands) are visible. While not problematic for some applications, this limitation substantially impedes the use of OST HMDs for all applications that augment the user itself rather than the environment. Prominent examples of such applications include virtual fitting rooms~\cite{pachoulakis2012augmented,srinivasan2017implementation}, learning of full body movements for dancing or personal training~\cite{anderson2013youmove}, or entertainment. Virtual models that could augment the user in aforementioned scenarios are shown in Figure~\ref{Figure:VRModels}.\\
In this work, we address this problem and propose a proof-of-principle system consisting of an OST HMD that is used together with a physical mirror to create an MR environment centered around the user of the HMD. Augmentation of the user is achieved indirectly by anchoring mirrored versions of virtual objects to the skeleton of the user's reflection that is tracked using an RGBD camera rigidly mounted on the HMD. To the best of our knowledge, this is the first effort in combining OST HMDs and mirrors to enable MR augmentation of reflections and, thus, the user. We believe this setup to bear enormous potential in all the scenarios above where large mirrors are already an integral part of the establishment, e.\,g. in department stores, fitness studios, and potentially at home.

\subsection{Related Work}
As the combination of HMDs and mirrors has not yet been studied, we limit the review of prior work to systems that either use physical or virtual mirrors to provide an MR environment for the user or use HMDs to directly augment the user.

\paragraph{Mirrors in MR}
Half-silvered mirrors, also referred to as beam-splitters, have been used in combination with bright display devices that are located on the side opposing the user~\cite{bimber2005spatial}. Ushida \textit{et al.}~\cite{ushida2002mirror} and Anderson \textit{et al.}~\cite{anderson2013youmove} use rear-mounted projectors that illuminate the back of large half-silvered mirrors. A more compact realization of this type of smart mirror is achieved by attaching a flat-panel LCD screen to the back of the half-silvered mirror~\cite{fujinami2005awaremirror,sato2009mr,besserer2016fitmirror,solosmirror}. These systems are thin and can be wall-mounted, yet, they require very large LCD displays if the complete mirror surface is desired for augmentation. These methods are comparably inexpensive but display the virtual content is not displayed in the focal plane of the reflection, making it impossible to focus on both the augmentations and the reflection at the same time. One way of addressing this problem is by using very large displays that are placed on the rear side of the mirror with a distance equal to one of the users to the mirror~\cite{hauntedmansion,holoflector}. Despite the obvious advantage of displaying virtual objects in the same focal plane as the reflection, the design has two considerable disadvantages: first, it requires a large configuration space and second, the user is required to maintain a fixed distance to the reflective surface. While these drawbacks may not be of particular importance in Walt Disney World's \emph{Haunted Mansion}~\cite{hauntedmansion}, it substantially limits the system's applicability in most real world scenarios where users move freely and space is constrained.\\ 
Other approaches replace real with virtual mirrors that are realized as video screens paired with optical cameras. Such systems have been used, e\,g., for cosmetic products~\cite{rahman2010augmented,javornik2016revealing} but also in the medical context for anatomy education~\cite{blum2012mirracle,bauer2015living,Bork2017}.\\
If the virtual content is to be displayed in relation to the user, as is the case for most aforementioned systems, tracking solutions are required. Sato \textit{et al.}~\cite{sato2009mr} use stereo RGB cameras to track a hand-held marker in 3D, while Rahman \textit{et al.}~\cite{rahman2010augmented} rely on a single IR camera to track an active IR emitters to estimate the distance of the user from the display. However, most approaches make use of human pose estimation algorithms that track the skeleton of the user over time~\cite{holoflector,anderson2013youmove,besserer2016fitmirror,Bork2017}. In this case, many configurations rely on the Microsoft Kinect solutions~\cite{zhang2012microsoft} as they are affordable and, in most cases, sufficiently accurate. 
It is worth mentioning that, contingent on their design, all of the aforementioned solutions are non-portable and unaffordable.

\paragraph{Self-Augmentation Using HMDs}
Compared to real or virtual mirror-based systems, designs based on HMDs do not necessarily require a prepared environment as the tracking sensors, such as the Leap Motion~\cite{weichert2013analysis}, can be integrated with the HMD and are thus located on the user rather than in its environment. Nevertheless, self-augmentation has not yet received much attention in the context of HMDs, particularly not when OST HMDs are considered.\\
Virtual reality (VR) headsets, such as the Oculus Rift (Oculus VR, Menlo Park, CA), employ concepts similar to self-augmentation in order to transfer the user's hands into the VR environment~\cite{Leap2017}. These applications are user focused and, in principle, allow for the augmentation of the tracked extremities with virtual content when the HMD is operated in video see-through mode.\\
Although in a very different context, namely for touch-less interaction, Lv et al.~\cite{lv2015touch} consider MR gaming environments in which the user's hands or feet are tracked to interact with the MR environment that is provided by Google Glass (Google, Mountain View, CA). 
Finally, the most relevant work was reported as a part of the LINDSAY Virtual Human project~\cite{tworek2013lindsay}. In their \emph{anARtomy} application, they use the Meta 1 OST HMD (Meta Company, San Mateo, CA) and track the user's hands using the HMD's optical camera and self-encoded markers that are attached to the front and back of both hands. The hands and forearms of the user are then augmented with anatomical models of the distal extremities~\cite{anartomy2015}.

\section{Methods}
\label{Methods}
\begin{figure*}[tb]
\centering
\includegraphics[width = 0.85\textwidth]{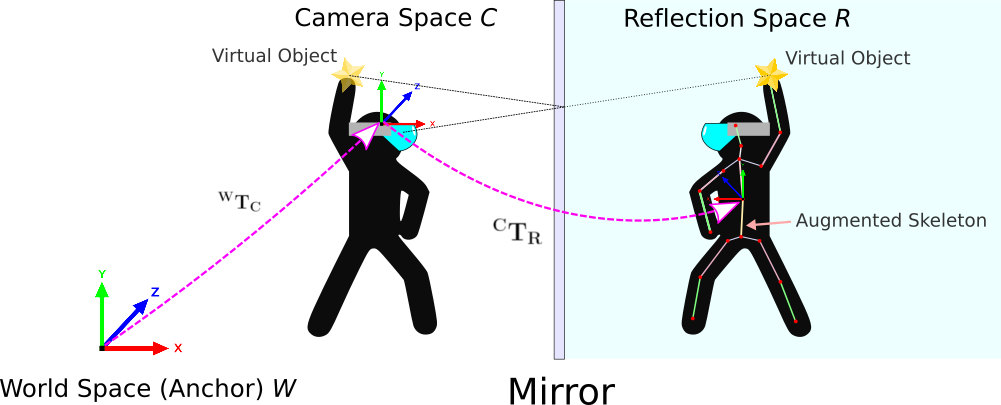}
\caption{Schematic of the working principle of the proposed system: the mirror creates a virtual image of the user in reflection space. An RGBD camera is used to track the skeleton of the user in the reflection space. Once the transformations $\trans{W}{C}$ (estimated using a SLAM-based algorithm) and $\trans{C}{R}$ (computed from the skeleton tracking information) are known, holograms can be anchored to the tracked skeleton points. We illustrate this relation by anchoring a star-shaped virtual object to the raised hand of the stick-figure in reflection space. In the MR environment provided to the user, the virtual object appears to augment its raised hand due to the reflection at the mirror surface indicated as thin black lines. Note that the coordinate system origins are offset to allow for a clearer visualization.}
\label{fig:schematic}
\end{figure*}

\subsection{System Overview}

\begin{figure*}
\centering
\includegraphics[width = 0.75\textwidth]{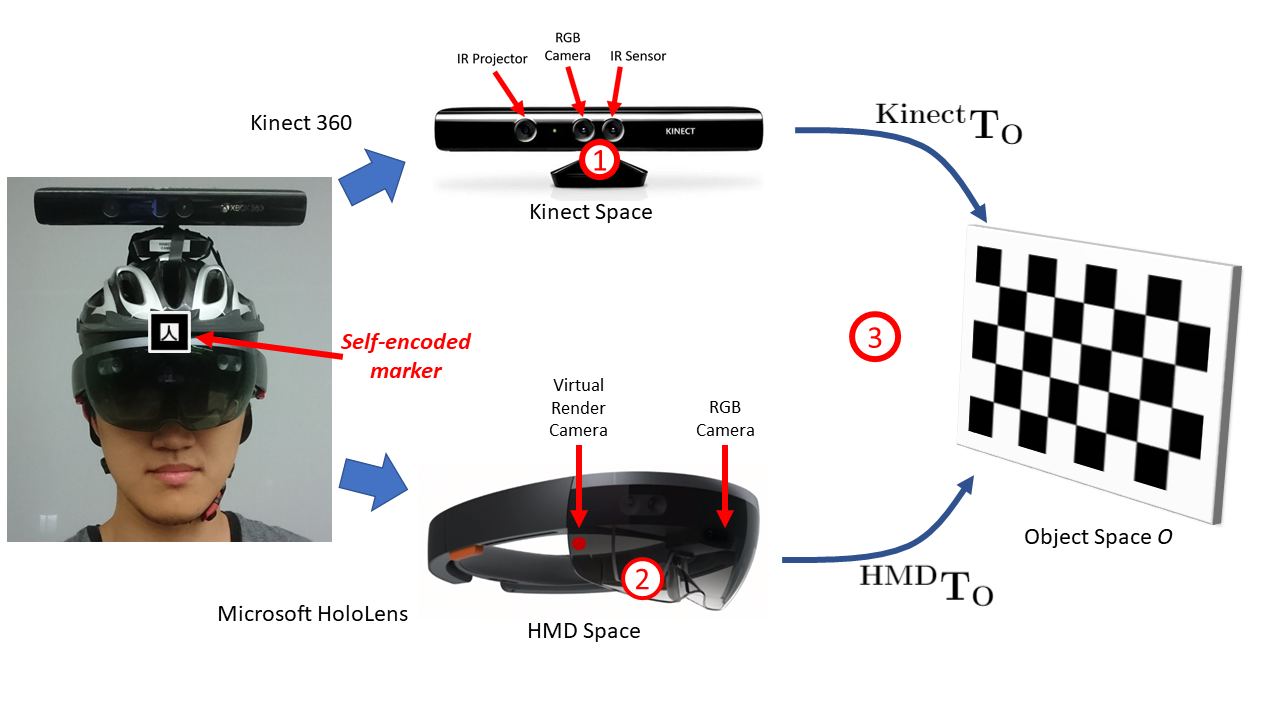}
\caption{Illustration of our prototype device (left) together with the required calibration steps. (1) Kinect internal: Calibration of the depth sensors to the RGB camera. (2) HoloLens internal: Calibration of the RGB camera to the virtual render camera. (3) Estimation of the extrinsic parameters $\trans{Kinect}{HMD}$ to calibrate the Kinect v1 to the virtual render camera of the HoloLens using a single ARToolkit marker.}
\label{fig:prototypeCalibration}
\end{figure*}

We propose an HMD-based system that enables self-augmentation of the user in quasi-unprepared environments; its working principle is illustrated schematically in Figure~\ref{fig:schematic}. 
In this context, quasi-unprepared refers to the fact that no display or motion capture system has to be installed and calibrated within the environment, yet, a sufficiently large mirror needs to be available. The proposed system comprises of
\begin{itemize}
  \item a device or software solution for human pose estimation, i.\,e. skeleton tracking (Section~\ref{subsubsec:skeleton}), and
  \item a self-encoded marker (Section~\ref{subsubsec:marker})
\end{itemize}
that are both tightly integrated with or rigidly attached to an OST HMD (Section~\ref{subsubsec:hmd}) and thus move with the user. We elaborate on design considerations and implementation details in the remainder of this manuscript and provide a picture of our proof-of-principle prototype in Figure~\ref{fig:prototypeCalibration}.

\subsection{Head mounted display}%
\label{subsubsec:hmd}
An OST HMD is an MR wearable display for the head. Virtual content is projected into the users eyes using either holographic~\cite{moon2014holographic} or fixed focus displays~\cite{birkfellner2003computer,Microsoft2017}. While holographic displays are able to render virtual content in the correct focal plane, stereoscopic fixed focus displays are more common and appropriate for several applications. Moreover, calibration of these devices is alleviated as the focal plane of the virtual render camera remains constant.\\
Within this work, we have selected the Microsoft's HoloLens as the foundation for our prototype. The HoloLens is a stereoscopic fixed focus OST HMD that is equipped with inertial and depth sensors, as well as a front facing RGB camera that combined enable accurate SLAM~\cite{Microsoft2017}. Apart from the RGB camera, all sensory output is unavailable. However, the HoloLens continuously provides an estimate of $\trans{W}{C}$ describing its pose with respect to the world coordinate system. The pose of the RGB camera can be derived from $\trans{W}{C}$, as the extrinsic parameters between the pose virtual render camera $\trans{W}{C}$ and the RGB camera on the HoloLens are provided by the manufacturer (see calibration step (2) in Figure~\ref{fig:prototypeCalibration}).\\
It is worth mentioning that for self-augmentation using a physical mirror as described here it is not imperative to know the pose of the HMD with respect to the environment. This is because the skeleton tracking in the reflection as described in Section~\ref{subsubsec:skeleton} is performed relative to the current pose of the HMD. However, knowledge of $\trans{W}{C}$ is beneficial if the virtual content is shared among multiple devices as described by Microsoft~\cite{microsoftSharing}. 

\subsection{Skeleton Tracking}%
\label{subsubsec:skeleton}
We augment the user by anchoring virtual content to its body. To this end, the pose of the user, commonly referred to as its skeleton, must be tracked in the reflection.
Skeleton tracking can be achieved in many different ways, however, the system considered here imposes some limitations as the device used for skeleton tracking a) must be tightly integrated or rigidly attached to the HMD such that it can be pre-calibrated to the HMD and b) must work on reflections. Both RGB and RGBD cameras fulfill these criteria~\cite{habert2015rgbdx} and are, therefore, the obvious choices. Considering that the HMD used here only allows access to the RGB camera feed, a skeleton tracking method based on monocular RGB images~\cite{li20143d,zhou2016sparseness} would be ideal, as it would allow a fully integrated prototype. Unfortunately, the execution of such methods on the HoloLens is not yet possible due to questionable robustness of these novel methods combined with hardware restrictions that would impede its execution in real-time. For our prototype described here, we decided to use an external RGBD sensor for skeleton tracking, namely the Kinect v1 (Kinect 360), that is rigidly attached to the HMD. This is achieved by attaching both the HoloLens and the Kinect v1 camera to a helmet as shown in Figure \ref{fig:prototypeCalibration}. While use of an additional, non-integrated sensor requires additional calibration steps (kindly refer to Figure~\ref{fig:prototypeCalibration}), the convenience and reliability of the skeleton tracking provided by the Microsoft Kinect SDK 1.8 outweighed the drawbacks for this implementation. Figure \ref{fig:prototypeCalibration} illustrates the prototype helmet and the calibration steps in more detail.

\paragraph{Skeleton Tracking Using the Kinect v1}%
The Microsoft Kinect cameras are consumer grade RGBD cameras that allow for image-based skeleton tracking of one or multiple persons in real-time~\cite{zhang2012microsoft}. 
It uses structured light to compute a depth image with a rather low resolution of $320\times240$ pixels in real-time. This depth image is then used as input to the Microsoft Kinect SDK 1.8 that estimates a 20-joint skeleton for every detected person. 
The Kinect v1 was not designed to be used on reflections; however, conventional mirrors have similar reflective properties in the infra-red and the visible spectrum~\cite{habert2015rgbdx}, suggesting that the Kinect will track the skeleton in the virtual image created by the mirror. The skeleton tracking algorithm has no mechanism to infer whether the current scene is a true or virtual, i.\,e. reflected, image. This introduces ambiguity for self-augmentation: if the skeleton is tracked in a reflected image, the association of skeleton joints with the left and right side of the body must be inverted to guarantee correct overlay. To resolve this ambiguity, a self-encoded marker is used (see Section~\ref{subsubsec:marker}).\\
It is worth mentioning that the Microsoft Kinect v2 is known to outperform its predecessor in all respects as it is using Time-of-Flight (ToF) rather than structured light but was found inadequate for our use-case. As the camera is rigidly attached to the HMD and, thus, the user's head it is constantly moving. The Kinect v2 has a constant and comparably long integration time of the ToF signal in order to increase its signal-to-noise ratio~\cite{lachat2015first,fursattel2016comparative}, that in the presence of perpetual motion leads to a substantially corrupted depth images impeding skeleton tracking.

\paragraph{Co-calibration}%
The Kinect v1 camera needs to be calibrated to the front facing RGB camera and, thus, the virtual render camera of the HMD. As indicated in Figure~\ref{fig:prototypeCalibration}, a total of three calibration steps are required:
\begin{enumerate}
\item Kinect v1 internal: Depth sensor to RGB camera\\
This calibration is achieved using a stereo-camera calibration using a checkerboard.
It is needed since the co-calibration in step 3 retrieves the extrinsic calibration of the RGB camera of the Kinect v1 with respect to the RGB camera of the HoloLens. 
\item HoloLens internal: Front-facing RGB camera to the virtual render camera\\
As mentioned in Section~\ref{subsubsec:hmd}, the extrinsic parameters of the virtual render camera acting as a display for the virtual content and the RGB camera are dissimilar. Fortunately, this calibration is readily provided by the HoloLens SDK.
\item Co-calibration: RGB camera of the Kinect to virtual render camera of the HoloLens\\
Once the individual devices are calibrated, co-calibration can be performed via standard stereo-calibration using either a checkerboard or a simple \emph{ARToolKit} marker. Stereo-calibration using checkerboard images yields more accurate results, but is time consuming. In the current prototype small changes in $\trans{Kinect}{HMD} = \trans{HMD}{O}^{-1}~\trans{Kinect}{O}$ cannot be excluded, we estimate $\trans{Kinect}{HMD}$ using an \emph{ARToolKit} marker that is detected and tracked in the RGB cameras both on the Kinect and the HoloLens. While this procedure is less accurate compared to traditional checkerboard stereo-calibration, it allows for re-calibration if the MR content appears displaced. 
\end{enumerate}

\subsubsection{Self-encoded Marker for Mirror Recognition}%
\label{subsubsec:marker}

As motivated in Section~\ref{subsubsec:skeleton}, the skeleton tracking algorithm works similarly on images of humans and their reflections suggesting that, from the human pose estimation alone, it is unclear whether the retrieved skeleton originates from a real person or the reflection of the user. This is problematic for self-augmentation, as mirrors are reversing such that the virtual content needs to be reverted as well to retain the true orientation. To this end, a self-encoded marker is rigidly attached to the front of the HMD as shown in Figure~\ref{fig:prototypeCalibration}. Then, detection and recognition of the self-encoded marker in the sensory feed of the HMD are tantamount with mirror detection.\\
In our prototype system we use an \emph{ARToolKit} marker that is glued to the visor of the HoloLens. The \emph{ARToolKit} software development kit (SDK)~\cite{kato1999marker} is used for recognition and tracking of the marker in the real-time video stream provided by the HoloLens' front-facing RGB camera.\\ 
The \emph{ARToolKit} SDK provides the 3D position of the detected marker with respect to the RGB camera. Note, that this position lies behind the reflective surface (denoted as reflection space as in Figure~\ref{fig:schematic}) as the marker detection cannot distinguish between real and virtual images. We use this ambiguity to compute the position of the reflective surface. The mirror is estimated as a plane located midway between the RGB camera center and the midpoint of the marker obtained from the \emph{ARToolKit} SDK. This line is simultaneously used as the normal of the estimated reflective plane.\\
Knowledge of the position and orientation of the reflective surface allows for the definition of a $4\times4$ involutive isometric transformation that describes a reflection with respect to the mirror plane~\cite{householder1958unitary}. This transformation, in turn, can be used to associate the skeleton points tracked in the reflection with their true position on the user. While this is not of immediate importance for the use-case reported here, it may prove useful in shared MR environments, particularly if $\trans{W}{C}$ is known.

\begin{figure}[tb!]
\centering
\includegraphics[width = 0.95\linewidth]{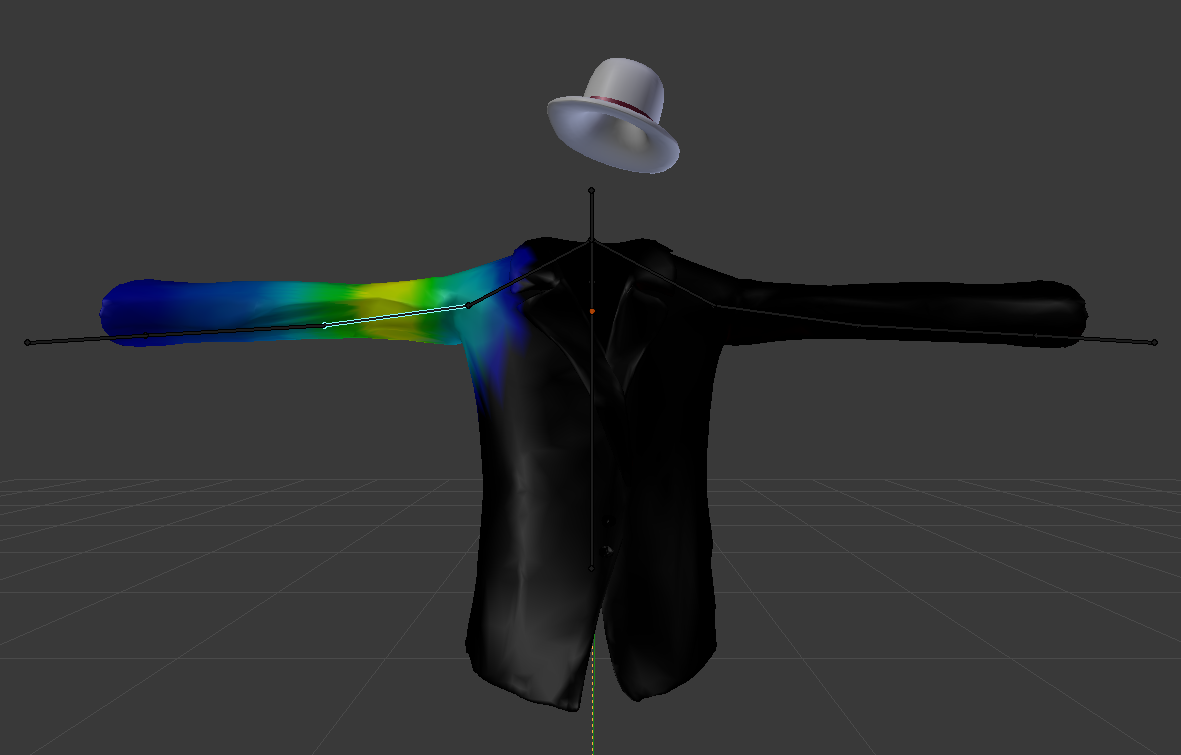}\\
(a) Model of a jacket with hat. Here the right upper sleeve is weighted and assigned to the upper arm bone of the armature.\\
\includegraphics[width = 0.95\linewidth]{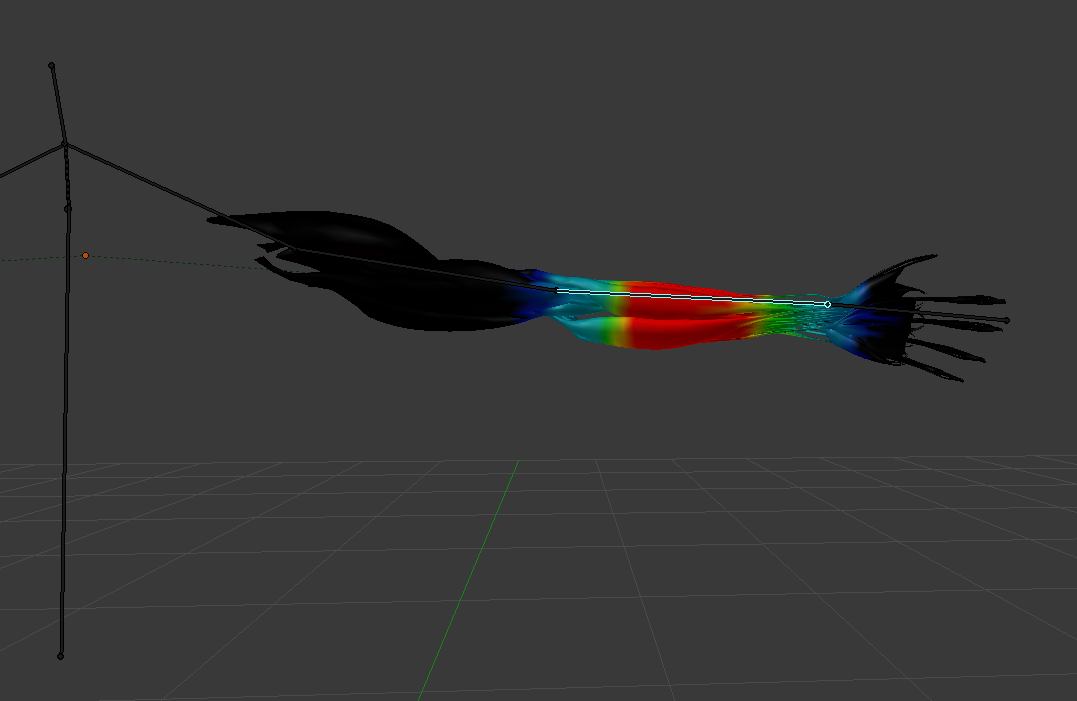}\\
(b) Musculary model of an upper extremity.Here the left forearm is weighted and assigned to the lower arm bone of the armature.
\caption{Vertex painting was done in the 3D modeling software Blender. The models have to be assigned to an armature which serves as an skeleton for deformations of the mesh. }
\label{Figure:vertexWeight}
\end{figure}

\begin{figure}[tb!]
\centering
\includegraphics[width = 0.95\linewidth]{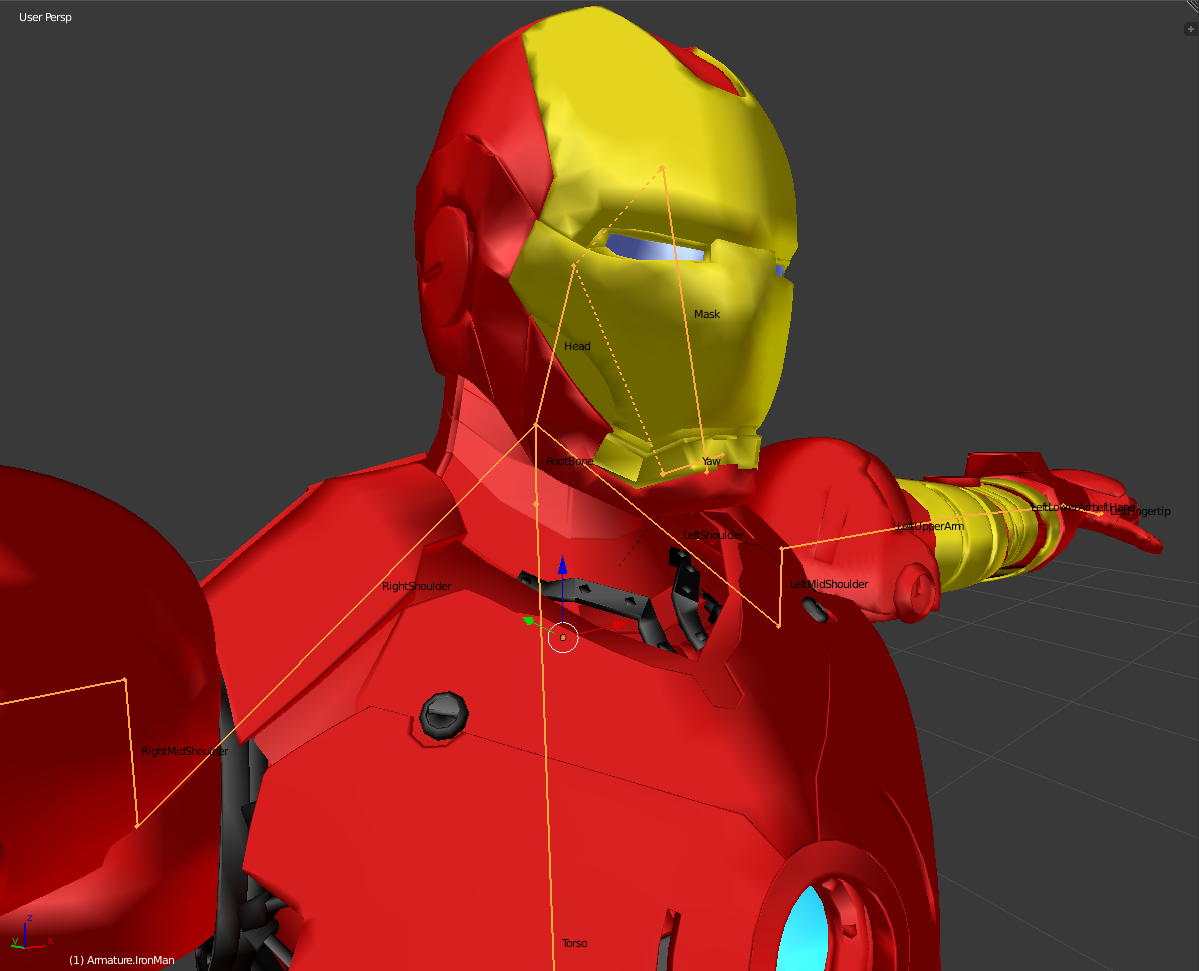}
\caption{Armature applied to the Ironman model. In the helmet area, two additional armature bones for the mask animation were added.}
\label{Figure:ironmanarmature}
\end{figure}

\paragraph{3D Models and Manipulation}%
Most currently existing work uses the skeleton to control a model with a constant scale, only using the rotational component of the skeleton bones. For a correct and visually pleasing overlay, however, the model has to be personalized to more accurately reflect the body shape of the detected person to promote immersion of the virtual model with the real scene. To this end, our system measures the shoulder distance and torso height and maps these distances onto the models to allow a perfect fit onto the user's body.\\
To allow for on-the-fly deformation of the 3D models, they have to be rigged as described as in \cite{Flavell2010}. That includes creating the armature and assigning vertices to the bones of the armature. The armature can be thought of a real skeleton to manipulate vertices which depend on these bones on the skeleton. In our case, we designed the armature of the models such that it resembles the skeleton of the user provided by the Kinect. Once the armature is defined, vertices of the model are assigned to specific bones. The assignment is not binary but continuous: a weight is computed for every vertex-bone pair that depends on the anticipated effect of bone transformations onto the displacement of the considered vertex. The assignment process is shown exemplary in Figure~\ref{Figure:vertexWeight}, where blue and red colors correspond to low and high weights, respectively.\\
The next step for the deformation is the transfer of the Kinect skeleton onto the armature of the model. This is accomplished by aligning bones of the tracked skeleton with the corresponding bone in the armature. 
Additionally, the rotation along the bone, especially for the supination and pronation rotation of the upper extremity is an important part for immersion. The Kinect v1 skeleton does not provide a sophisticated hand tracking by itself, however, the cross product of the upper arm and lower arm facing towards the torso provides a good looking and reasonable facing direction for each bone of the arm.\\
The final and one of the most important steps for the self-augmentation, is the positioning of the model itself in space such that it overlays with the virtual image of the user created by the mirror. A single joint is selected to serve as the anchor joint that strictly follows a certain 3D point of the Kinect skeleton in space. This becomes necessary as all other verices are displaced with respect to this anchor joint to enable personalization. 
In our prototype system, we identified the middle shoulder joint on the height of the larynx as the most stable choice and used it in all experiments.\\
Detailed animation of the models is done in the Unity engine that powers our HoloLens app. This is beneficial, as it allows us to directly use all animation resources provided by Unity for our models. In our demonstrations, only the Ironman suit is animated. This is achieved by adding additional, artificial bones that allow for opening and closing of the ventail. This is shown in Figure~\ref{Figure:ironmanarmature}.\\
The preceding considerations hold true for animation of full body models but also for models that are limited to a particular body part. Without loss of generality, we limit the models considered here and shown in Figure~\ref{Figure:VRModels} to the upper torso and head. This has a practical reason: Full body models do not fit into the narrow field of view of the HoloLens such that full body models would not promote an improved MR experience. On the contrary, the increased complexity of full body models leads to performance decrease due to the limited compute power of the HoloLens.

\subsection{Experiments and Demonstrations}%
We seek to assess the appropriateness of our prototype system and demonstrate its applicability to self-augmentation. To this end, we first quantitatively assess the calibration accuracy of the Kinect v1 to the HoloLens, and the degradation of the IR pattern used for depth perception that is associated with imperfect reflectivity of the mirror. 

\subsubsection{Calibration Repeatability and Accuracy}%
Using the AR-marker for co-calibrating the HoloLens and Kinect is a very fast alternative to retrieve the extrinsic calibration but may be of low quality. Therefore, we evaluate the repeatability of the co-calibration by computing the Euclidean distance from post-calibration translations and rotational deviation compared to the initial calibration. To this end, we collect 30 measurements over one minute after the initial calibration and evaluate the error. This process is repeated 5 times where a new calibration is estimated each time.\\
We also determine an upper bound on the co-calibration accuracy using an alternative method that uses a conventional checkerboard with the size of $12\times7$ and a square side length of $4.5$\,cm. We simultaneously take snapshots from both devices and create image pairs. Those serve as input for a checkerboard detection algorithm such that the corner points are known and the re-projection error can be calculated.

\subsubsection{Depth Signal Degradation due to Reflection}%
Depending on their coating, mirrors exhibit different reflective properties for light in the visible and IR spectrum. In general, the reflectivity in the IR spectrum is lower~\cite{habert2015rgbdx} suggesting that the intensity of the IR pattern used for depth image estimation is lower in reflections compared to real scenes. This degradation can affect the effective detection range and depth estimation performance of the Kinect.\\
To quantify the degradation, we compare the intensity of the measured IR pattern when a) looking at a person standing in front of the Kinect at a particular distance, and b) looking at the reflection of the user standing at exactly half the respective distance from a mirror. The experiments are performed under artificial light. Moreover, in both experiments the person wore the same clothes to guarantee the same scattering behavior of the IR light during the experiment. We report the average IR image intensity as a function of the distance from the Kinect camera.

\subsubsection{Demonstrations}%
Since the system is highly portable, it enables MR self-augmentation in various different scenarios and for numerous applications. We demonstrate the feasibility of MR self-augmentation using physical mirror in three different scenarios where large mirrors are commonly used such that the proposed system could find immediate application: virtual fitting rooms, anatomy learning and personal fitness, and entertainment.

\paragraph{Virtual Fitting Room}
The most straightforward application of the proposed system is the virtual fitting room. This scenario was considered in previous work but required prepared environments~\cite{pachoulakis2012augmented,srinivasan2017implementation} to track the user and display the virtual content. The virtual fitting room experience can be achieved by our system, that only requires the proposed HMD and a body-sized mirror to be installed in the environment of the user. The virtual fitting room MR environment is demonstrated using the jacket and hat model shown in Figure~\ref{Figure:VRModels}.

\paragraph{Anatomy Learning and Personal Fitness}
Similar to the MagicMirror system~\cite{blum2012mirracle,Bork2017}, the system finds applications in health-care education, where anatomical structures such as bones are overlaid at the correct position providing an interactive learning environment. For personal fitness applications, our prototype may act as a low-cost motion capture system~\cite{fern2012biomechanical} that provides real-time information on the user's joint dynamics. This information can be used to a) counting the number of repetitions in weight lifting exercise and b) inform the user on incorrectly performed movements via MR overlay of the correct motion sequence. We believe that this real-time feedback could be of particular benefit in rehabilitation.  

\paragraph{Entertainment and Gaming}
Self-augmentation with 3D models of fantasy universes, such as the animated Ironman suit, demonstrates the application of the proposed system for entertainment applications. While we do not explicitly demonstrate the use of our system for gaming, we believe its deployment for games such as the ones introduced in conjunction with the \emph{EyeToy}'s Kung-Foo~\cite{kungfoo,marks2010eyetoy} to be relatively straightforward.

\section{Results}
\label{Results}
\subsection{Quantitative Assessment}

\subsubsection{Calibration Repeatability and Accuracy}


The mean co-calibration error achieved when the simple marker-based setup was used to estimate the rigid transformation between HoloLens and Kinect is $2.78 \pm 2.28$\,cm while the median error is $2.17$\,cm. More importantly the rotational error is $1.35 \pm 0.86$\,degrees and median error of $1.15$\,degrees.\\
The alternative method for calibration using a checkerboard yields an re-projection error of $1.66$\,pixels.

\subsubsection{Depth Signal Degradation due to Reflection}
\begin{figure}[tb!]
\centering
\includegraphics[width = 0.95\linewidth]{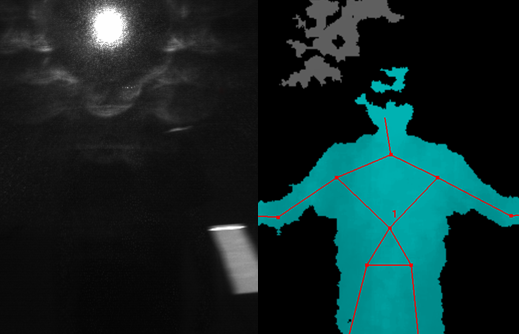}
\caption{Left: Raw IR image. Right: Processed and segmented depth image. Direct reflection of the IR pattern saturates the infrared sensor and obstructs the system from seeing the head in the depth image.}
\label{Figure:irArtefact}
\end{figure}

\begin{figure}[tb!]
\centering
\includegraphics[width = 0.95\linewidth]{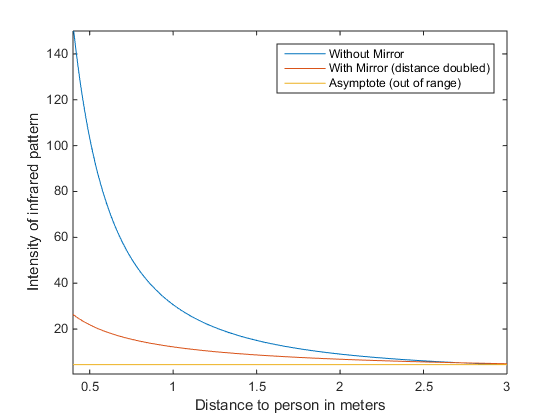}
\caption{Curve comparison of the intensity of the IR pattern as a function of the distance to the person standing in front of the camera. The asymptote indicates the intensity threshold below which skeleton tracking cannot be performed reliably.}
\label{Figure:irCurve}
\end{figure}

From Figure~\ref{Figure:irArtefact} it becomes apparent that the user is visible in the IR image of the Kinect acquired from the reflection. However, the IR emitter is also imaged leading to substantial artifact. Yet, the Microsoft Kinect SDK 1.8 is able to reliably detect and track the skeleton of the upper torso.\\
Intensity curves of the IR pattern at various target distances from the sensor are shown in Figure~\ref{Figure:irCurve}. Distances reported for the reflection consider the distance of the Kinect to the virtual image (double the distance to the mirror). Moreover, we show the asymptote that corresponds to the case where no IR signal can be measured. While the IR pattern intensity is considerably lower in reflected compared to real scenes, the intensity is still well above the threshold used for out-of-range areas indicated by the asymptote if the target of interest is sufficiently close to the reflective surface. Based on these results, we position the user at a distance of approximately $1.00$\,m to the mirror to ensure functionality of the skeleton tracking.   

\subsection{Demonstrations}

\begin{figure}[tb!]
\centering
\includegraphics[width=0.95\linewidth]{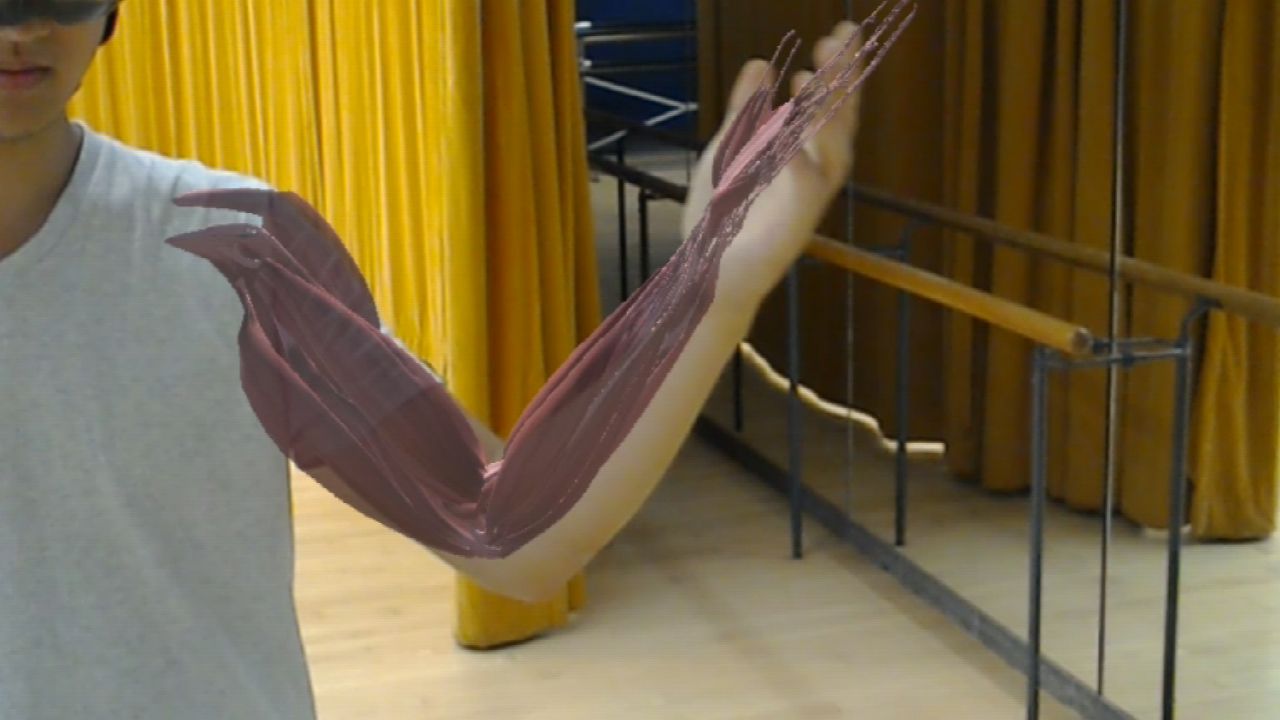}\\
(a) Arm muscles
\includegraphics[width = 0.95\linewidth]{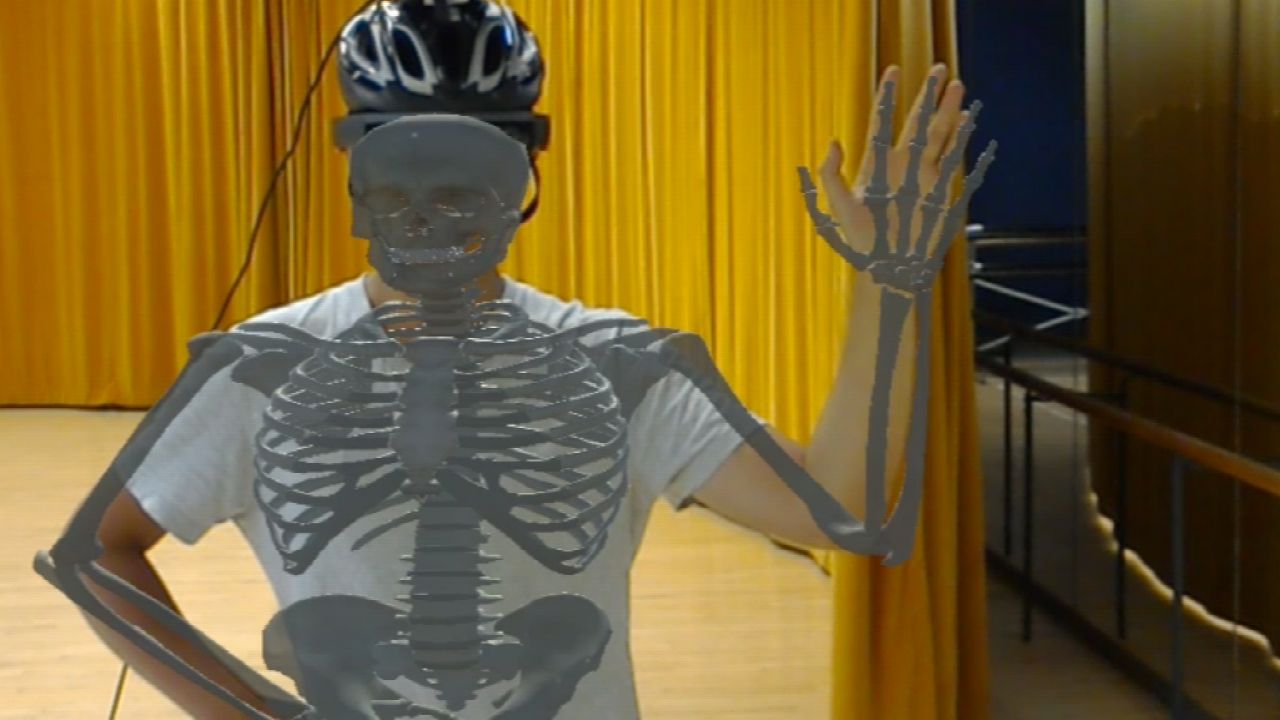}\\
(b) Skeleton\\
\includegraphics[width=0.95\linewidth]{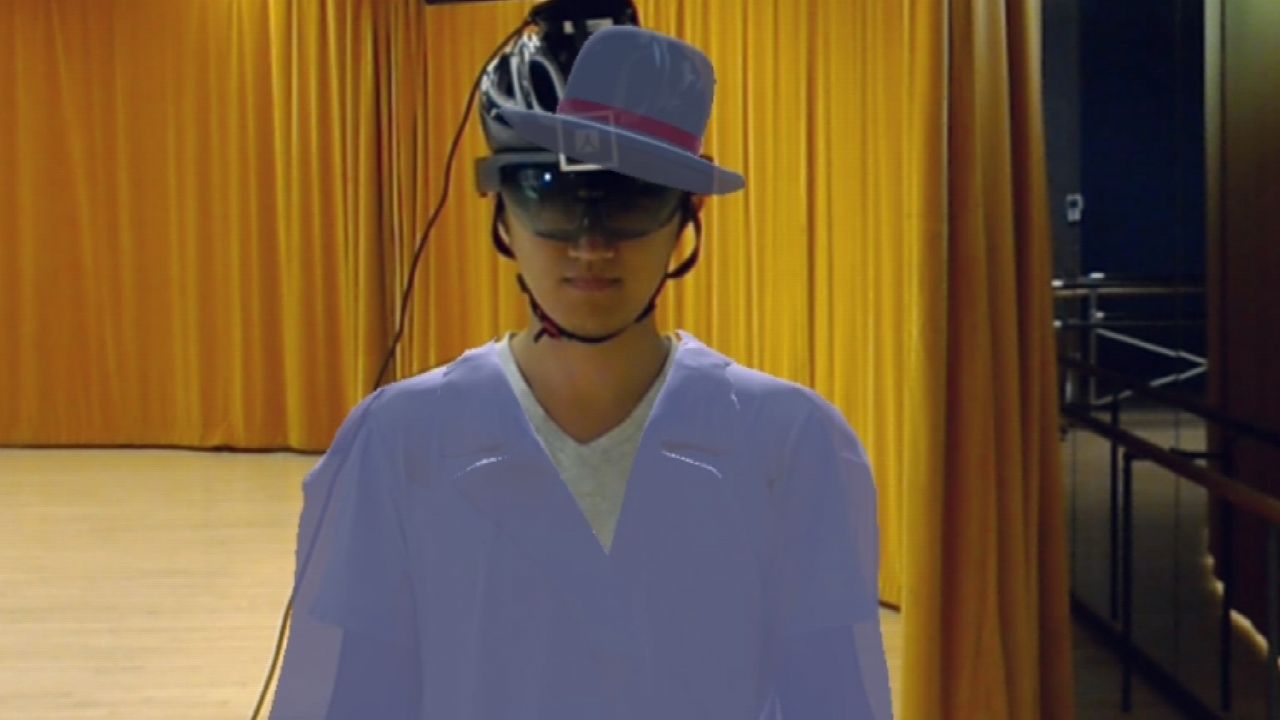}\\
(c) Jacket and hat
\includegraphics[width=0.95\linewidth]{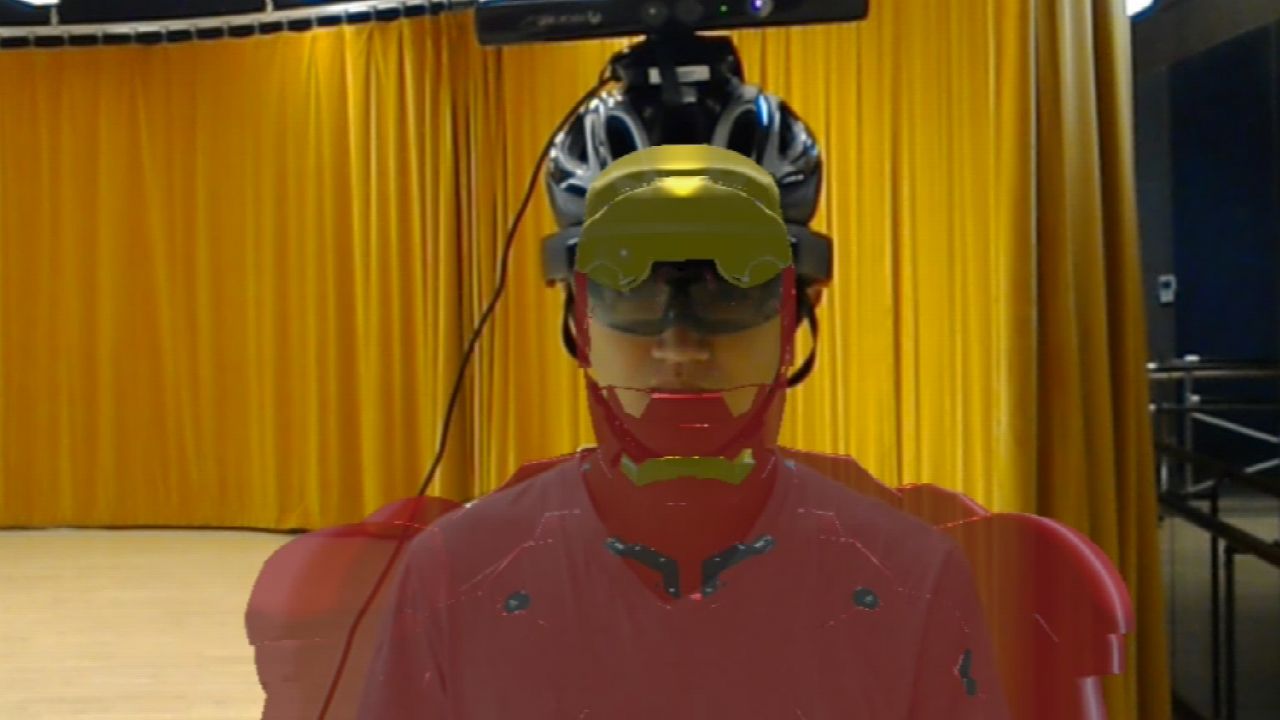}\\
(d) Hero suit
\caption{Self-augmentation using the models. These images were recorded using the HMD directly and strongly resemble the perception of the user. Due to the skeleton tracking, the user can freely move in front of the mirror while the augmentation with the 3D model is maintained and adapted to match the current pose of the user's upper extremities.}
\label{Figure:skeletonar}
\end{figure}

Self-augmentation with the models were perceived well. Skeleton tracking as provided by the Kinect v1 and co-calibration performed sufficiently well to enable credible immersion. The rendered models exhibit high contrast and opaqueness which can be appreciated in Figure~\ref{Figure:skeletonar}. Dark colors possess a low visibility since they are equivalent to the transparency factor due to the nature of the additive color display mechanism of the OST HMD. \\
The communication between the Kinect and the HoloLens on a single dedicated wireless local network allows for reliable real-time communication with a delay of approximately $20$\,ms.

\section{Discussion}
\label{Discussion}

In summary, our proof-of-principle prototype system was able to provide convincing evidence that self-augmentation using OST HMDs and reflections from physical mirrors is possible and potentially useful in several applications. However, the proposed method and device could substantially benefit from future work.\\
The most obvious drawback of the current prototype is that skeleton tracking in the reflection is not fully integrated with the HMD and may, therefore, suffer from de-calibration over time. In such cases, use of the outdated calibration results in an incorrect overlay of the virtual content with the reflection. While our results suggest that the accuracy of our simple online calibration using a simple \emph{ARToolKit} marker is acceptable in most cases, use of the HMD integrated sensors would allow for very accurate offline calibration without the need for frequent re-calibration. The authors believe that, as previously mentioned in the manuscript, human pose estimation methods that operate on monocular RGB images~\cite{li20143d,zhou2016sparseness} would lend themselves well for the considered use-case. In addition, methods that operate on RGB rather than IR images could increase the flexibility of the proposed system. The reason is that the reflectivity of conventional mirrors is much higher for the visible compared to the IR spectrum. This, in turn, mitigates the effect observed in Figure~\ref{Figure:irCurve}, where limited reflectivity in the IR spectrum substantially limited the distance of the user from the mirror. However, as the proposed system is able to detect the mirror using the self-encoded marker, the user can be directed towards the mirror in cases when skeleton tracking fails.\\
It is worth mentioning that the proposed system is directly applicable to tracking and augmentation of other persons in the user's environment. This emphasized the importance of reflection detection, as models used for self-augmentation need to be flipped for accurate overlay. In our prototype, a self-encoded marker was used to detect reflections that performed well, however, novel methods based on face detection and recognition could be deployed that would not require additional markers and may even work if the face is partly occluded by the visor of the HMD~\cite{singh2017disguised}.\\
During our experiments we noted misalignments of the deformed model and the user that arise from de-synchronization. Considering the low latency of our data transfer network of about $20$\,ms, the observed lag results from the limited hardware resources of the HoloLens. The processing power of the HMD is not sufficient to accurately align and deform the high resolution 3D models to the user's skeleton provided by the Kinect.\\
Improvements of OST HMDs will immediately benefit the proposed system, making us confident that the proposed concept of combining OST HMDs with physical mirrors for self-augmentation will be found useful. Applications in tele-rehabilitation and personal training may be of particular interest, as real-time feedback overlaid directly with the user can be a key advantage.

\section{Conclusion}
\label{Conclusion}
In this paper, we proposed the concept of optical see-through head-mounted display (OST HMD) based self-augmentation and a prototype implementation. The system anchors virtual content displayed in the HMD to the reflection of the user that is generated by a physical mirror. 
Based on skeleton tracking in the reflection, we demonstrate the capabilities of the system in augmenting the user with humanoid models. 
Compared to existing systems, our system requires only an OST HMD equipped with an RGBD camera making it highly portable. Therefore, it can be used in any facility with pre-installed body-sized mirrors including, but not limited to, applications in health-care, entertainment, and personal fitness.\\
The promising performance of our prototype device motivates further refinement of the system in order to explore its capabilities in tele-rehabilitation, where we believe the system could be of particular importance as it combines real-time motion capture with mixed reality feedback directly overlaid with the user.

\bibliographystyle{abbrv-doi-hyperref}
\bibliography{main}

\end{document}